\documentclass[twocolumn,twocolappendix,appendixfloats]{aastex6}

\usepackage{subfigure}
\usepackage{threeparttable}
\usepackage{multirow}
\usepackage{remreset}
\usepackage{CJKutf8}

\collaborationName{Friends of AASTeX}
\shorttitle{Gradual transition from Type S to Type N BAL}
\shortauthors{Lu \& Lin}
\begin{document}


\title{Narrow absorption lines complex. III. Gradual transition from Type S to Type N Broad Absorption Line}


\author{
\begin{CJK*}{UTF8}{gbsn}
Wei-Jian Lu (陆伟坚)\altaffilmark{1} and Ying-Ru Lin (林樱如)\altaffilmark{2}
\end{CJK*}
}
\affil{School of Information Engineering, Baise University, Baise 533000, China}

\altaffiltext{1}{E-mail: william\_lo@qq.com (W-J L)}
\altaffiltext{2}{E-mail: yingru\_lin@qq.com (Y-R L)}



\begin{abstract}
We study the relationship between the broad absorption lines (BALs) that can be decomposed into multiple narrow absorption lines and those that cannot (hereafter Type N and Type S BAL, respectively), based on the analysis of three BAL systems (systems A, B, and C) in two-epoch spectra of quasar SDSS J113009.40+495247.9 (hereafter J1130+4952). As the velocity decreases (from systems A to C), these three BAL systems show a gradual transition from Type S to Type N BAL, and their equivalent widths (EWs) and profile shapes vary in a regular way. We ascribe the absorption line variability in J1130+4952 to the ionization change as a response to the fluctuation of the ionizing continuum based on several factors: (1) coordinated EW strengthening over a wide range in systems A and B, (2) the system B shows an obvious change in Si\,{\footnotesize IV} but no significant change in C\,{\footnotesize IV} BAL, and (3) asynchronized variability between the continuum and absorption lines. Based on the analysis of the variation mechanism, location, ionization state, and structure of systems A, B, and C, we hold the view that the Type S and Type N BALs in J1130+4952 probe the same clumped outflow, with the Type S BALs originating from the inner part of the outflow with a relatively higher ionization state, smaller column density and more clumpy structures, while the Type N BALs originate from the outer part of the outflow with relatively lower ionization state, larger column density, and fewer clumpy structures.

\end{abstract}




\section{Introduction} \label{sec:intro}

Blueshifted absorption lines in rest-frame ultraviolet (UV) spectra of quasars are powerful tools in probing outflows and the feedback process to the host galaxies (e.g., \citealp{DiMatteo2005, Hopkins2008,Fabian2012}). According to the width of absorption profiles, absorption lines can be divided into three categories: broad absorption lines (BALs, absorption widths of at least 2000 $\rm km\,s^{-1}$, e.g., \citealp{Weymann1991}), narrow absorption lines (NALs, {full width at
half maximum (FWHM) narrower than 500} $\rm km\,s^{-1}$, e.g., \citealp{Misawa2007}), and mini-broad absorption lines (mini-BALs, absorption widths between those of BALs and NALs, e.g., \citealp{Hamann2004}). {According to previous research, the outflow NALs in quasar spectra are more common than the other two absorption line types. Their detection rate in C\,{\footnotesize IV} ions in quasar spectra are roughly $\sim$45\% for NAL, $\sim$5\% for mini-BALs, and $\sim$20\% for BALs (\citealp{Hamann2012}).}

Studies of the physical relationship between these three observed line types can provide valuable information for understanding the quasar outflows. They could represent different stages of the quasar evolution (e.g., \citealp{Farrah2007,Hamann2008}), or different sight lines of the quasar inclination (e.g., \citealp{Murray1995,Elvis2000,Proga2000}). In each case, research has increasingly suggested that these three types of absorption lines in quasar spectra may not belong to essentially separate classes, instead they may probe the same general outflow phenomenon. This conclusion is supported by at least the following research results. First, transitions of one type of outflow line into another type have been reported (\citealp{Leighly2009,Hall2011,Filiz2013,Rodriguez2013,Rogerson2016,Moravec2017}). Second, different types of lines have been simultaneously detected in individual quasars (e.g., \citealp{Moe2009,Misawa2014a,Misawa2014b,Moravec2017}). Third, correlations between the variability of absorption lines and the continuum are proved in systematic studies on different types of lines (\citealp{Lu2017,Chen2018a,Chen2018b} for NAL; \citealp{Wang2015,He2017,LLQ2018,Lu2018saturation,Vivek2019} for BAL). {These correlations indicate that the ionization change (IC) is the dominant mechanism for UV absorption line variations, although they cannot completely rule out other variation mechanisms such as transverse motion (TM) of absorbing gas across sight lines.}

Moreover, it has recently been reported that some BALs (Type N BALs) can be easily decomposed into multiple NALs (\citealp{Lu2018complex1,Lu2018complex2,Lu2018saturation,Stathopoulos2019}), indicating some of the BALs and intrinsic NALs are essentially the same. However, there are also some BALs (Type S BALs) that have a relatively smooth trough that cannot be decomposed into separated NALs. This raises a new question: what is the physical relationship between Type N BAL and Type S BAL? 

In this paper, we report the study of a quasar, J113009.40+495247.9 ($z=2.085$; \citealp{Paris2017}, hereafter J1130+4952), from the Sloan Digital Sky Survey (SDSS), which shows both Type N and Type S BALs in its two-epoch spectra. This interesting discovery may offer clues to understanding the physical connection between these two BAL types. The structure of the paper is as follows. Section \ref{sec:spectro} describes the spectral analysis. Section \ref{sec:disscu} provides the discussions on the variation mechanism, location, ionization state, and structure of the absorption systems. Section \ref{sec:Conclusion} includes a summary. Throughout the paper, a $\Lambda$CDM cosmology with parameters $H_0=70\,\rm km\,s^{-1}\,Mpc^{-1}$, $\Omega_{\rm M}=0.3$, and $\Omega_{\Lambda}=0.7$  is adopted.

\begin{table*}[h]
    \centering
\caption{Measurements of Si\,{\footnotesize IV} and C\,{\footnotesize IV} absorption lines \label{tab.1}}
\begin{tabular}{lcccccccc} 
\hline 
\hline 
Species & $z_{\rm abs}$ & Velocity & \multicolumn2c{MJD:52642} & \multicolumn2c{MJD:56412} & Fractional  &Note\\
\cline{4-7}
 & & &EW &FWHM &EW &FWHM &EW &    \\
 & &($\rm km~s^{-1}$) & (\AA) & ($\rm km~s^{-1}$) & (\AA) & ($\rm km~s^{-1}$) & Variation & \\
\hline
Si\,{\footnotesize IV}$\lambda$1393 	&	 { 2.0061 } 	&	 { 7770 } 	&	 $ 1.01 \pm 0.12 $ 	&	370	&	 $ 1.10 \pm 0.05 $ 	&	370	&	 $ 0.09 \pm 0.127 $ 	&	 { Si1 } \\
Si\,{\footnotesize IV}$\lambda$1402 	&	2.0062	&	7757 	&	 $ 0.73 \pm 0.17 $ 	&	368	&	 $ 0.83 \pm 0.06 $ 	&	351	&	 $ 0.13 \pm 0.243 $ 	&	 ... \\
Si\,{\footnotesize IV}$\lambda$1393 	&	 { 2.0147 } 	&	 { 6907 } 	&	 $ 0.92 \pm 0.13 $ 	&	369	&	 $ 0.99 \pm 0.06 $ 	&	369	&	 $ 0.07 \pm 0.154 $ 	&	 { Si2 } \\
Si\,{\footnotesize IV}$\lambda$1402 	&	2.0149	&	6885 	&	 $ 0.50 \pm 0.30 $ 	&	417	&	 $ 0.50 \pm 0.12 $ 	&	372	&	 $ 0.00 \pm 0.646 $ 	&	 ... \\
Si\,{\footnotesize IV}$\lambda$1393 	&	 { 2.0221 } 	&	 { 6173 } 	&	 $ 0.99 \pm 0.14 $ 	&	419	&	 $ 1.32 \pm 0.05 $ 	&	419	&	 $ 0.29 \pm 0.143 $ 	&	 { Si3 } \\
Si\,{\footnotesize IV}$\lambda$1402 	&	2.0219	&	6193 	&	 $ 1.09 \pm 0.12 $ 	&	405	&	 $ 0.93 \pm 0.05 $ 	&	333	&	 $ -0.16 \pm 0.122 $ 	&	 ... \\
Si\,{\footnotesize IV}$\lambda$1393 	&	 { 2.0331 } 	&	 { 5088 } 	&	 $ 0.53 \pm 0.18 $ 	&	340	&	 $ 0.53 \pm 0.07 $ 	&	285	&	 $ 0.00 \pm 0.364 $ 	&	 { Si4 } \\
Si\,{\footnotesize IV}$\lambda$1402 	&	2.0329	&	5108 	&	 $ 0.31 \pm 0.21 $ 	&	248	&	 $ 0.34 \pm 0.07 $ 	&	198	&	 $ 0.09 \pm 0.707 $ 	&	 ... \\
Si\,{\footnotesize IV}$\lambda$1393 	&	 { 2.0404 } 	&	 { 4367 } 	&	 $ 0.31 \pm 0.18 $ 	&	246	&	 $ 0.31 \pm 0.11 $ 	&	266	&	 $ 0.00 \pm 0.680 $ 	&	 { Si5 } \\
Si\,{\footnotesize IV}$\lambda$1402 	&	2.0402	&	4387 	&	 $ 0.20 \pm 0.29 $ 	&	248	&	 $ 0.32 \pm 0.09 $ 	&	248	&	 $ 0.46 \pm 1.398 $ 	&	 ... \\
Si\,{\footnotesize IV}$\lambda$1393 	&	 { 2.0437 } 	&	 { 4039 } 	&	 $ 0.16 \pm 0.38 $ 	&	249	&	 $ 0.33 \pm 0.09 $ 	&	249	&	 $ 0.69 \pm 2.103 $ 	&	 { Si6 } \\
Si\,{\footnotesize IV}$\lambda$1402 	&	2.0439	&	4019 	&	 $ 0.14 \pm 0.34 $ 	&	198	&	 $ 0.20 \pm 0.10 $ 	&	198	&	 $ 0.35 \pm 2.402 $ 	&	 ... \\
C\,{\footnotesize IV} BAL 	&	 ... 	&	 $ -22718 \sim -18588^{\rm a} $ 	&	 $ 3.14 \pm 0.40 $ 	&	 4129$^{\rm b}$ 	&	 $ 9.10 \pm 0.19 $ 	&	 4129$^{\rm b}$ 	&	 $ 0.97 \pm 0.100 $ 	&	 system A \\
Si\,{\footnotesize IV}BAL 	&	 ... 	&	 $ -22718 \sim -18784^{\rm a} $ 	&	 $ 0.25 \pm 0.49 $ 	&	 3934$^{\rm b}$ 	&	 $ 2.82 \pm 0.24 $ 	&	 3934$^{\rm b}$ 	&	 $ 1.67 \pm 0.581 $ 	&	...\\
C\,{\footnotesize IV}BAL 	&	 ... 	&	 $-16746 \sim -10478^{\rm a} $ 	&	 $ 22.14 \pm 0.42 $ 	&	 6268$^{\rm b}$ 	&	 $ 22.02 \pm 0.20 $ 	&	 6268$^{\rm b}$ 	&	 $ -0.01 \pm 0.021 $ 	&	 system B \\
Si\,{\footnotesize IV}BAL 	&	 ... 	&	 $-16746 \sim -9781^{\rm a} $ 	&	 $ 8.28 \pm 0.61 $ 	&	 6965$^{\rm b}$ 	&	 $ 12.35 \pm 0.28 $ 	&	 6965$^{\rm b}$ 	&	 $ 0.39 \pm 0.075 $ 	&	...\\
C\,{\footnotesize IV}BAL 	&	 ... 	&	 $-9478 \sim -3152^{\rm a} $ 	&	 $ 22.35 \pm 0.35 $ 	&	 6325$^{\rm b}$ 	&	 $ 22.00 \pm 0.18 $ 	&	 6325$^{\rm b}$ 	&	 $ -0.02 \pm 0.018 $ 	&	 system C \\
Si\,{\footnotesize IV}BAL 	&	 ... 	&	 $ -9478 \sim -1889^{\rm a} $ 	&	 $ 7.86 \pm 0.58 $ 	&	 7588$^{\rm b}$ 	&	 $ 8.52 \pm 0.29 $ 	&	 7588$^{\rm b}$ 	&	 $ 0.08 \pm 0.081 $ 	&	... \\

\hline 
\end{tabular}
\begin{tablenotes}
\footnotesize
\item$^{\rm a}$Velocity range of the BAL troughs with respect to the emission rest frame.
\item$^{\rm b}$Total width calculated from edge-to-edge of the BAL trough.
\end{tablenotes}
\end{table*}


\begin{figure*}
\includegraphics[width=2\columnwidth]{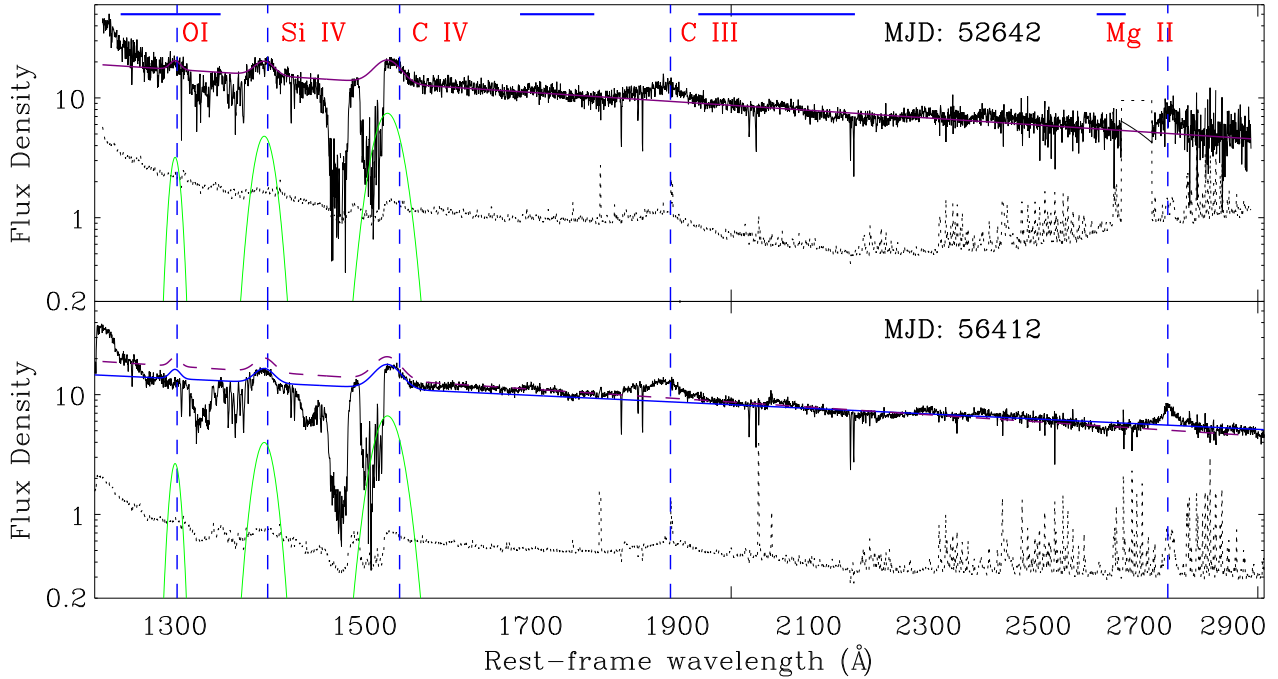}
\caption{Final pseudo-continuum fits (the purple and blue smooth curves) for the two-epoch spectra of quasar J1130+4952. The flux density is in units of $\rm 10^{-17}~erg~s^{-1}~cm^{-2}$. The blue vertical dashed lines mark out the main emission lines. The blue horizontal bars on top of the upper panel represent the relatively line-free regions chosen for the power-law fits. {The purple dashed line in the bottom panel is the final pseudo-continuum fit for the MJD
52642 spectrum.} The dotted lines near the bottom of each panel are the formal 1$\sigma$ errors. Green Gaussian profiles in the bottom of each panel are fits for the emission lines. Both the longitudinal and transverse axes are logarithmic.}  \label{fig.1}
\end{figure*}

\begin{figure}
\includegraphics[width=1\columnwidth]{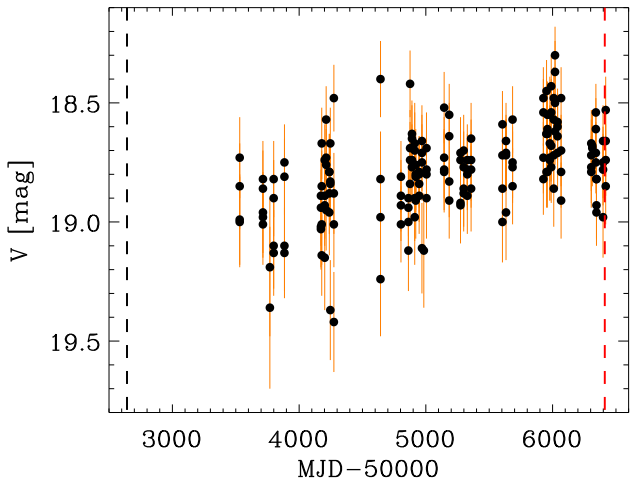}
\caption{Light curve of J1130+4952 from the CRTS. Vertical black and red dashed lines indicate the timing of the SDSS and BOSS spectral observations, respectively.}
\label{fig.2}
\end{figure}
\begin{figure*}
\includegraphics[width=2\columnwidth]{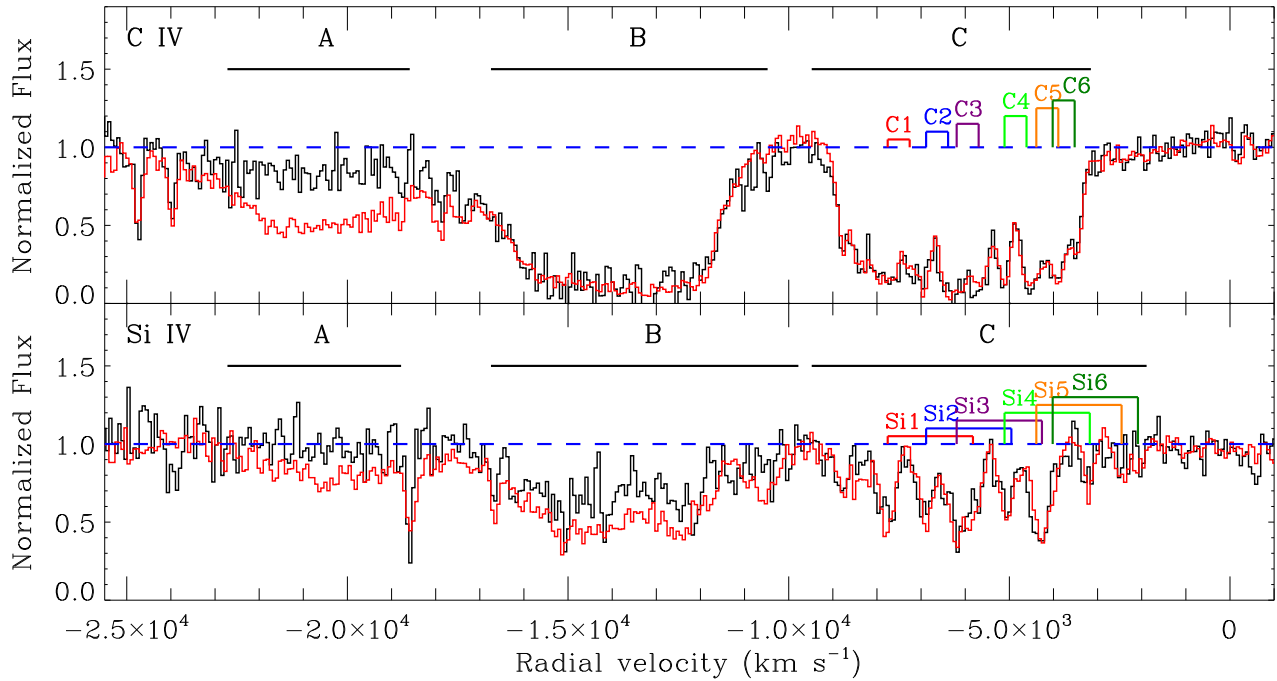}
\caption{Portions of the pseudo-continuum normalized spectra of J1130+4952. The black and red lines represent the MJD 52,642 and 56,412 spectra, respectively. Normalized fluxes are plotted vs. the radial velocity of the strongest member (1548.195 and 1393.755 for C\,{\footnotesize IV} and Si\,{\footnotesize IV} absorption doublets respectively), with respect to the emission-line redshift of 2.085. The red, blue, purple, green, orange, and dark green vertical lines mark out the six identified NAL systems. Black horizontal lines on the top indicate the studied BAL systems. }
\label{fig.3}
\end{figure*}
\begin{figure}
\includegraphics[width=1\columnwidth]{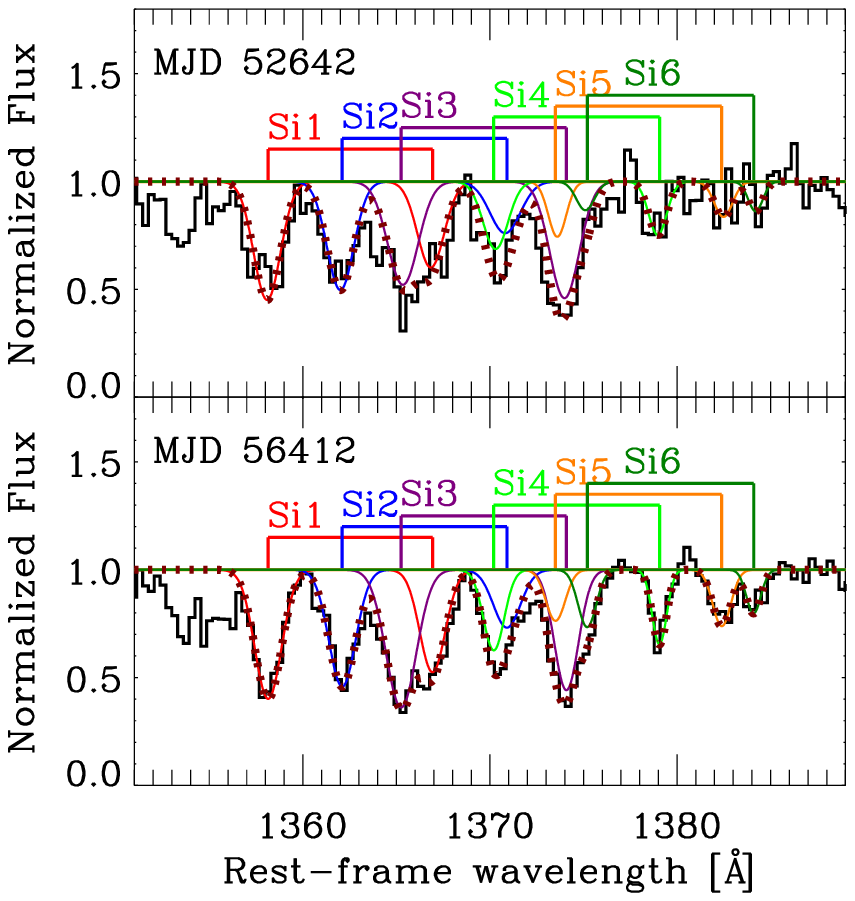}
\caption{Identified NALs within the Si\,{\footnotesize IV} BAL trough of system C in J1130+4952. The top and bottom spectra are snippets from the MJD 52,642 and 56,412 normalized spectra, respectively. The brown dotted lines represent the total fit model.} \label{fig.4}
\end{figure}


\section{SPECTROSCOPIC ANALYSIS and results} \label{sec:spectro} 
J1130+4952 has been observed twice by the SDSS and {SDSS-III's Baryon Oscillation Spectroscopic Survey (SDSS-BOSS; \citealp{York2000,Dawson2013})}, on MJD 52,642 and 56,412, with a wavelength region of $\sim$3800--9200, and $\sim$3600--10000\,\AA, respectively. We downloaded the two-epoch spectra of J1130+4952 from SDSS data release 14 (DR14; \citealp{Abolfathi2018}) where the flux calibration has been improved. These two spectra from DR14 have resolutions of $\sim$1500 at 3800\,\AA~and $\sim$2500 at 9000\,\AA. Their median signal-to-noise ratio (S/N) are 10.42 (for MJD 52,642 spectrum) and 17.78 (for MJD 56,412 spectrum) per pixel, respectively.

The procedure of spectroscopic analysis in this paper is similar with our previous works (\citealp{Lu2018complex1,Lu2018complex2,Lu2018saturation}). In short, the power-law continua were fitted iteratively based on a few wavelength regions (1250--1350, 1700--1800, 1950--2200, and 2650--2710\,\AA~ in the rest frame; which is defined by \citealp{Gibson2009a}). Several Gaussian profiles are applied to fit the emission lines. The final pseudo-continuum is a combination of the power-law continuum fit and the fits to the emission lines (Figure \ref{fig.1}). 
{In order to further confirm the variation of the quasar continuum, we utilized the $V$-band photometric observations of J1130+4952 from the Catalina Real-Time Transient Survey (CRTS; \citealp{Drake2009}). CRTS is a large time-domain optical survey with a coverage area of $\sim33,000\,\rm deg^2$ and a baseline of 8 yr, with each target exposing approximately 250 times per year. The survey was performed using unfiltered light and then calibrated to a $V$-band zeropoint.  The light curve of J1130+4952 from CRTS is shown in Figure \ref{fig.2}.}

Then we identified absorption lines in the spectra that normalized using the pseudo-continua. {As shown in Figure \ref{fig.3}, we have detected three BAL systems in J1130+4952 (see also Table \ref{tab.1}). Moreover, as the velocity decreases (from systems A to C), we found a gradual transition between these three systems in both their equivalent widths (EWs) and profile shapes. 
In EWs, both C\,{\footnotesize IV} and Si\,{\footnotesize IV} BALs of system A show large variation, while both C\,{\footnotesize IV} and Si\,{\footnotesize IV} BALs of system C show no significant variation\footnotemark[1]. System B is in a state of transition between systems A and C, whose Si\,{\footnotesize IV} BAL shows obvious variation but no significant change in C\,{\footnotesize IV} (hereafter Phenomenon I).
In profile shapes, both C\,{\footnotesize IV} and Si\,{\footnotesize IV} BALs of system A are almost undetectable in the first-epoch SDSS spectrum, but the C\,{\footnotesize IV} BAL in the second-epoch spectrum has a relatively smooth semicircular profile that cannot be decomposed into NALs, making it a Type S BAL; the Si\,{\footnotesize IV} BAL of system C can be directly resolved into several NALs, and the corresponding NALs of the same system in its C\,{\footnotesize IV} BAL are severely blended, but the obvious outline of multiple NAL troughs still be shown in bottom, making it a Type N BAL; system B is in a state of transition between Type N and Type S BALs, whose C\,{\footnotesize IV} BAL has a square-bottomed profile with a smooth tail on both sides of the trough, while its Si\,{\footnotesize IV} BAL are also cannot be decomposed into NALs but a few NAL troughs on the bottom can still be distinguished. }

\footnotetext[1]{In fact, some Si\,{\scriptsize IV} NALs in system C (for example, systems Si3 and Si6) may show time variation, although they have large errors due to the low S/N of the spectra.}

{Note that, in the velocity range from $\sim-17,000\rm \,km\,s^{-1}$ to $-18,000\rm \,km\,s^{-1}$ in the top panel of Figure \ref{fig.3}, the continuum flux of the BOSS spectrum is absorbed more than 10\%. According to the definition of BAL (\citealp{Weymann1991}), the C\,{\footnotesize IV} in systems A and B could be contained in a single BAL trough. However, as mentioned above, the C\,{\footnotesize IV} in systems A and B show obvious differences in both their EWs and profile shapes, we thus separately analyzed them in this paper.}

We adopted six pairs of Gaussian functions to fit the NALs within the Si\,{\footnotesize IV} BAL trough in system C (Figure \ref{fig.4}). Note that the numbers of the absorption systems we have identified are only the lower limit because of the low resolution of the spectra, as well as the blending of the NALs. For example, there is absorption in the blue side of the Si\,{\footnotesize IV} BAL in system C, which means that six Gaussian profiles cannot completely match the whole Si\,{\footnotesize IV} BAL. Then we measured the velocity, EW, FWHM, and fractional EW variation of the NALs based on these Gaussian fits. The methods of calculating the EWs and the corresponding errors of the BALs are the same as those in Lu \& Lin (\citeyear{Lu2018complex1}; see their Equations (2) and (3)), and the equations for the fractional-EW-variation and the corresponding error calculations are the same as those in Lu et al. (\citeyear{LLQ2018}; see their Equations (2) and (3)). Measured values of the NALs and BALs are listed in Table \ref{tab.1}.

\section{DISCUSSION} \label{sec:disscu}
Results in Section \ref{sec:spectro} provide clues to understanding the physical relationship between systems A, B, and C. 
One of the most vital points is the gradual transition between these three systems. 
The simultaneous detection of systems A, B, and C, as well as their gradual transition in EW and profile shape described above indicate a close physical relationship between these three systems. Below, we will discuss the physical relationships between systems A, B, and C from the following aspects: variation mechanism, location, ionization state, and structure.

\subsection{{Variation mechanism}} \label{sec:Variation mechanism} 
Variability of a BAL may be caused by a TM and/or IC scenario of an outflow. We hold the opinion that the latter mechanism is responsible for the BAL variability in J1130+4952. Our reasons are as follows.

First, between the two-epoch observations, EW strengthening over the whole BAL trough is detected in the C\,{\footnotesize IV} and Si\,{\footnotesize IV} BALs in system A and the Si\,{\footnotesize IV} BAL in system B. Synchronous changes between multiple NAL/BAL systems (e.g., \citealp{Hamann2011,Chen2015,Wang2015,McGraw2017}) or over a large continuous range of velocity interval in a BAL trough (e.g., \citealp{Grier2015,McGraw2017,McGraw2018}) could serve as strong evidence that supports the IC scenario to explain variation in absorption lines. It is difficult to explain the phenomenon of coordinated absorption line change through the TM scenario because it requires coordinated motions of numerous distinct outflow structures (\citealp{Misawa2005,Hamann2011}). 

Second, Phenomenon I, shown in system B, disfavours the TM scenario, because the TM scenario would likely change the trough profile of a C\,{\footnotesize IV} BAL that is highly optically thick (e.g., \citealp{Misawa2014a,Capellupo2014,Wang2015,McGraw2018,Lu2018saturation}). In \citealp{Lu2018saturation}, a moderate anticorrelation between the fractional changes of Si\,{\footnotesize IV} BALs and the UV continuum was confirmed in 74 quasars that show Phenomenon I, revealing the ubiquitous effect of the UV continuum variability on Phenomenon I. Phenomenon I also indicates a physical process, strengthening in recombination-driven column density with no significant EW variation, in the saturated C\,{\footnotesize IV} BAL (\citealp{Lu2018saturation}). 

Third, asynchronized variability has been identified between the continuum and absorption lines. Anticorrelations between the changes of the ionizing continuum and BALs have been proven (\citealp{LLQ2018,Lu2018saturation,Vivek2019}), which supports the IC scenario as the primary driver of the variability of broad absorption troughs. Therefore, the continuum variation can account for the variability in BAL to some extent. In J1130+4952, when the BALs show coordinated strengthening (see Figure \ref{fig.3}) between the two-epoch observations, the power-law continuum appears to have a fractional weakening\footnotemark[2] of $-0.066\pm0.0032$ (see Figure \ref{fig.1}). This phenomenon is consistent with the anticorrelation in the existing studies. 

\footnotetext[2]{We choose the observed wavelength range of $4000–9000$\,\AA~ to obtain the flux of the power-law continuum and the corresponding error is calculated by the error propagation method.}

Based on the above analysis, we attribute the absorption line variability in J1130+4952 to the IC scenario, which is actually the response to the ionizing continuum variability. 
{According to the photoionization simulations, with an increasing ionization parameter ($U$), an EW of C\,{\footnotesize IV} or Si\,{\footnotesize IV} would rise first, then arrive at a peak, and decrease at the end (e.g., figure 2a of \citealp{Hamann1997}). In J1130+4952, asynchronized variability between the continuum and systems A and B indicates that these BALs are at a relatively high ionization state. }
Besides, {w}e attribute the lack of variation in C\,{\footnotesize IV} BAL of system B, as well
as in both C\,{\footnotesize IV} and Si\,{\footnotesize IV} BALs of system C, to saturation, which can be inferred from several aspects: (1) the Phenomenon I shown in system B, (2) the square-bottomed profiles in C\,{\footnotesize IV} BALs of systems B and C, and (3)$\sim$1:1 doublet ratios in some of the resolved Si\,{\footnotesize IV} NALs of system C. However, these saturated lines do not reach zero intensity, which indicates partial coverage of these outflow clouds to the background light source. Considering the variation of continuum as the mechanism driving the absorption line variability in J1130+4952, we speculate that the recombination-driven column density strengthening may occur in these saturated BALs, although no significant EW variation is detected. Finally, based on the gradual transition of both EW variation and the saturation between systems A, B, and C, we obtained the column densities $N_{\rm HA}\textless N_{\rm HB}\textless N_{\rm HC}$.

\subsection{{Locations}} \label{sec:Locations} 
{Under the assumption that faster outflow is closer to the central engine and considering the observational evidence that the velocities between systems A, B, and C show a gradual transition, we can make an initial inference that the relationship between the  radial distances of the absorbers responsible for systems A, B, and C are $r_{\rm A}\textless r_{\rm B}\textless r_{\rm C}$.}

{We can make further estimations of their distances from the central engine in two different ways. On the one hand, as shown in Figure \ref{fig.1}, system C is positioned near the center of the broad emission lines (BELs), and the absorption depths of both its Si\,{\footnotesize IV} and C\,{\footnotesize IV} BALs are deeper than the corresponding BELs, while the system A is positioned at higher velocity with no strong emission lines at its wavelengths, which suggests that the absorbers producing system C cover both the continuum source and broad emission-line region (BELR), while the absorbers producing system A cover only the continuum source. Thus, their distances ($r$) from the flux source should be that beyond BELR for absorbers producing system C.}

{On the other hand, if we simply assume that (1) the ions we measure are the dominant ionization stages and (2) the gas is in an ionization equilibrium state, then we can limit the electron density and the distance of the absorber from the ionizing photon source via using the variability timescale as the upper limit of the recombination time (e.g., \citealp{Narayanan2004,Grier2015}). In the case of  J1130+4952, the C\,{\footnotesize IV} varies between two-epoch observations over a timescale of $\Delta t_{\rm obs}=3770$\,days (i.e., an rest-frame timescale of $\Delta t_{\rm rest}=1222$\,days), so we can estimate a lower limit value for the electron density of the absorption gas as $n_e\textgreater3.383\times10^3\,\rm cm^{-3}$, as well as an upper limit on its distance with respect to the center source as $r\textless2648\,\rm pc$. By visually checking Figure \ref{fig.2},  we find that the weakening of the V band magnitude was actually happened after MJD$\sim$56,000. If we take the MJD of 56,000 as the start time of the recombination process, then the timescale for this process is $\Delta t_{\rm obs}=412$\,days (i.e., $\Delta t_{\rm rest} = 134$\,days), which corresponding to a minimum electron density of $n_e\textgreater 3.095\times10^4\,\rm cm^{-3}$, and a maximum distance from the center source of $r\textless875\,\rm pc$.}

\subsection{{Ionization state}} \label{sec:Ionization state} 
The distance of absorption gas to the center can,  to some extent, reveal the ionization parameter $U$, which is defined as 
\begin{equation}
	    U=\frac{1}{4\pi r^2n_Hc}Q,
    \label{eq:quadratic 1}
\end{equation}
where $Q$ is the emission rate of hydrogen ionization photons. Since they are illuminated by the same background source, systems A, B, and C have the same parameter $Q$. According to Equation (\ref{eq:quadratic 1}), a closer distance from the center indicates a higher ionization parameter $U$. Thus, in J1130+4952, we got $U_{\rm A}\textgreater U_{\rm B}\textgreater U_{\rm C}$. 

Under the situation in which the IC scenario is the mechanism responsible for the BAL variation in J1130+4952, the relationship of the ionization state between systems A, B, and C can, to some extent, be inferred from the gradual EW variation of systems A, B, and C through the photoionization equilibrium analysis (\citealp{Osterbrock2006,Wang2015}). 
As described in Section \ref{sec:spectro}, although illuminated by the same background source, BALs with larger velocity show larger fractional variation. In other words, systems A, B, and C show different response in EW to the same continuum variation. Assuming that the shape of the ionization continuum remains the same and the ionization is controlled by C$^{4+}$, then the number ratio of C$^{3+}$ to C$^{4+}$ can be approximated with (see the detailed calculation process in section 5.4 of \citealp{Wang2015})
\begin{equation}
	    \frac{n_{\rm C^{3+}}}{n_{\rm C^{4+}}}\propto U^{-1},
    \label{eq:quadratic 2}
\end{equation}
{where the $n_{\rm C^{3+}}$ and $n_{\rm C^{4+}}$ represent the numbers of ions of C$^{3+}$ and C$^{4+}$ respectively.} If C$^{5+}$ is the dominant species, then $n_{\rm C^{3+}}\propto U^{-2}$; if C$^{6+}$ is the dominant one, then $n_{\rm C^{3+}}\propto U^{-3}$. A similar analysis can be  performed for Si$^{3+}$. From these calculations we can find that gases with higher ionization states are more sensitive to the ionizing continuum variation. Thus, in J1130+4952, we determine $U_{\rm A}\textgreater U_{\rm B}\textgreater U_{\rm C}$, which is consistent with the deduction from the discussion above.

\subsection{{Outflow structure and profile shape}} \label{sec:Outflow structure and profile shape} 
We think that the phenomenon of BAL consisting of NALs, as shown in system C, reveals the clumpy structure of BAL outflow. On the contrary, system A have relatively smooth trough that cannot be decomposed into distinct NALs. Does this mean that system A has a different outflow from system C, such as a smooth and homogeneous flow? This conjecture can be overturned when considering both their partial covering characteristics (Section \ref{sec:Variation mechanism}) and their distances relative to the BELR (Section \ref{sec:Locations}). The situation that there are no strong emission lines at the absorption line wavelength  of system A indicates that its partial covering pertains to the quasar continuum but not the BELR. This situation of partial covering requires small outflow absorbing structures, owning to {the spatial scale of} the UV continuum source is expected to be $\textless 0.01$\,pc across (\citealp{Hamann2011}). In other words, the outflow clouds responsible for system A also have clumpy structures instead of a smooth flow. Such outflow structures might resemble the schematic of inhomogeneous partial coverage (\citealp{Hamann2001,deKool2002,Hamann2004,Arav2005,Sabra2005}).

Because there is a common origin of clumped flow for systems A and C, then what is/are the physical reason(s) that led to the different profile shape between them? The origin of clumpy outflow structures for system A indicates that they are also consist of a superposition of NAL features, yet it cannot be resolved into distinct NALs; it even has no outline of the NAL features at its bottom (see Figure \ref{fig.3}). We think the most likely explanation for this is that the NAL features within system A are more severely blended than system C, which would require a large number of NALs considering their large velocity width (Table \ref{tab.1}). Based on the above inferences and the gradual transition of the profile shapes between systems A, B, and C, we can make a further inference that the relationship between the density of clumpy structures responsible for systems A, B, and C are  $\rho_{\rm A}\textgreater\rho_{\rm B}\textgreater\rho_{\rm C}$.

Note that a larger density of clumpy structures does not necessarily equate to a higher column density of specific absorption transition(s). Instead, system A in J1030+4952 has a lower column density (less saturation) than system C in transitions of both C\,{\footnotesize IV} and Si\,{\footnotesize IV} (Section \ref{sec:Variation mechanism}). This is reasonable when considering that the clumpy structures producing system A have closer distances to the background light source as well as higher degrees of ionization (Section \ref{sec:Ionization state}) than system C, {which is located beyond the BELR but has an upper limit of $\sim875\,\rm pc$ (Section \ref{sec:Locations}). }

There is other observational evidence that could be used to back up the above assertion. Although the C\,{\footnotesize IV} troughs of both systems B and C suffer from saturation, system B has tailing transition regions on both sides of the trough, while the two sides of the C\,{\footnotesize IV} trough of system C are straight up and observed to steepen. We think this is because the edges of system C only consist of single (or a few) NAL(s) with relatively lower ionization state(s) and larger EW(s), while the tail of system B is blended with a large number of NALs with higher ionization state(s) and smaller EW(s), which is also consistent with the above assertion. 

\subsection{{Implication}} \label{sec:Implication} 
Considering the common origin of a single outflow, we point out that it may be an evolutionary connection between systems A, B, and C in J1130+4952. In other words, the kinematic shift of system A that gradually varies to that of C may be expected in the follow-up observations. The deceleration of quasar winds seems to be expected in galactic feedback models (e.g., \citealp{Silk1998,DiMatteo2005,Fabian2012}), in which the interaction between outflows/winds and environmental material in the host galaxies is a crucial element. If it is indeed an evolutionary connection between systems A, B, and C, then their evolution time should be longer than a few years (rest-frame time), because no significant BAL deceleration is detected between the two-epoch spectra of J1130+4952 during rest-frame timescales of 3.35\,yr. Moreover, general BAL acceleration or deceleration have been proven to be rare based on a large sample of multiobserved SDSS quasars (\citealp{Grier2016}; rest-frame timescales of their sample range from 2.71 to 5.49 \,yr). 

\section{{Conclusion}} \label{sec:Conclusion} 
Based on the above results and discussions about the systems A, B, and C in J1130+4952, we can pry into the physical relationship between Type S and Type N BALs. The gradual transition of the spectra signatures and physical properties from system A to system C actually represents a gradual transition from Type S BAL to Type N BAL. From J1130+4952, we can find that the outflow clouds responsible for Type S and Type N BALs have both similarities and specialities. On the one hand, both of them originate from the same general clumped outflow, and their gradual transition also indicates that there is no clear boundary between them. On the other hand, the Type S BAL may originate from the inner part of the outflow with relatively higher ionization state, smaller column density, and more clumpy structures, while the Type N BALs originate from outer part of the outflow with a relatively lower ionization state, larger column density, and fewer clumpy structures.

\acknowledgments
We would like to acknowledge the anonymous reviewer for useful comments, which improved the paper.  

Funding for the Sloan Digital Sky Survey IV was
provided by the Alfred P. Sloan Foundation, the U.S.
Department of Energy Office of Science, and the Participating
Institutions. SDSS-IV acknowledges support and resources
from the Center for High-Performance Computing at the
University of Utah. The SDSS website is \url{http://www.sdss.org/}.

SDSS-IV is managed by the Astrophysical Research
Consortium for the Participating Institutions of the SDSS
Collaboration including the Brazilian Participation Group, the
Carnegie Institution for Science, Carnegie Mellon University,
the Chilean Participation Group, the French Participation
Group, Harvard-Smithsonian Center for Astrophysics, Instituto
de Astrofísica de Canarias, The Johns Hopkins University,
Kavli Institute for the Physics and Mathematics of the Universe
(IPMU)/University of Tokyo, Lawrence Berkeley National
Laboratory, Leibniz Institut für Astrophysik Potsdam (AIP),
Max-Planck-Institut für Astronomie (MPIA Heidelberg),
Max-Planck-Institut für Astrophysik (MPA Garching), MaxPlanck-Institut für Extraterrestrische Physik (MPE), National
Astronomical Observatories of China, New Mexico State
University, New York University, University of Notre Dame,
Observatário Nacional/MCTI, The Ohio State University,
Pennsylvania State University, Shanghai Astronomical Observatory, United Kingdom Participation Group, Universidad
Nacional Autónoma de México, University of Arizona,
University of Colorado Boulder, University of Oxford,
University of Portsmouth, University of Utah, University of
Virginia, University of Washington, University of Wisconsin,
Vanderbilt University, and Yale University.

\bibliographystyle{aasjournal}
\bibliography{NALvsBAL} 

\begin{thebibliography}{}
\expandafter\ifx\csname natexlab\endcsname\relax\def\natexlab#1{#1}\fi

\bibitem[{{Abolfathi} {et~al.}(2018){Abolfathi}, {Aguado}, {Aguilar}, {Allende
  Prieto}, {Almeida}, {Tasnim Ananna}, {Anders}, {Anderson}, {Andrews},
  {Anguiano}, \& et~al.}]{Abolfathi2018}
{Abolfathi}, B., {Aguado}, D.~S., {Aguilar}, G., {et~al.} 2018, \apjs, 235, 42

\bibitem[{{Arav} {et~al.}(2005){Arav}, {Kaastra}, {Kriss}, {Korista}, {Gabel},
  \& {Proga}}]{Arav2005}
{Arav}, N., {Kaastra}, J., {Kriss}, G.~A., {et~al.} 2005, \apj, 620, 665

\bibitem[{{Capellupo} {et~al.}(2014){Capellupo}, {Hamann}, \&
  {Barlow}}]{Capellupo2014}
{Capellupo}, D.~M., {Hamann}, F., \& {Barlow}, T.~A. 2014, \mnras, 444, 1893

\bibitem[{{Chen} {et~al.}(2018{\natexlab{a}}){Chen}, {Pang}, {He}, \&
  {Huang}}]{Chen2018a}
{Chen}, Z.-F., {Pang}, T.-T., {He}, B., \& {Huang}, Y. 2018{\natexlab{a}},
  \apjs, 236, 39

\bibitem[{{Chen} \& {Qin}(2015)}]{Chen2015}
{Chen}, Z.-F., \& {Qin}, Y.-P. 2015, \apj, 799, 63

\bibitem[{{Chen} {et~al.}(2018{\natexlab{b}}){Chen}, {Yao}, {Pang}, {Yi}, {Lu},
  {Liu}, {Nong}, {Liang}, {Liang}, {Ma}, {Wu}, {Gan}, \& {Zou}}]{Chen2018b}
{Chen}, Z.-F., {Yao}, M., {Pang}, T.-T., {et~al.} 2018{\natexlab{b}}, \apjs,
  239, 23

\bibitem[{{Dawson} {et~al.}(2013){Dawson}, {Schlegel}, {Ahn}, {Anderson},
  {Aubourg}, {Bailey}, {Barkhouser}, {Bautista}, {Beifiori}, {Berlind},
  {Bhardwaj}, {Bizyaev}, {Blake}, {Blanton}, {Blomqvist}, {Bolton}, {Borde},
  {Bovy}, {Brandt}, {Brewington}, {Brinkmann}, {Brown}, {Brownstein}, {Bundy},
  {Busca}, {Carithers}, {Carnero}, {Carr}, {Chen}, {Comparat}, {Connolly},
  {Cope}, {Croft}, {Cuesta}, {da Costa}, {Davenport}, {Delubac}, {de Putter},
  {Dhital}, {Ealet}, {Ebelke}, {Eisenstein}, {Escoffier}, {Fan}, {Filiz Ak},
  {Finley}, {Font-Ribera}, {G{\'e}nova-Santos}, {Gunn}, {Guo}, {Haggard},
  {Hall}, {Hamilton}, {Harris}, {Harris}, {Ho}, {Hogg}, {Holder}, {Honscheid},
  {Huehnerhoff}, {Jordan}, {Jordan}, {Kauffmann}, {Kazin}, {Kirkby}, {Klaene},
  {Kneib}, {Le Goff}, {Lee}, {Long}, {Loomis}, {Lundgren}, {Lupton}, {Maia},
  {Makler}, {Malanushenko}, {Malanushenko}, {Mandelbaum}, {Manera}, {Maraston},
  {Margala}, {Masters}, {McBride}, {McDonald}, {McGreer}, {McMahon}, {Mena},
  {Miralda-Escud{\'e}}, {Montero-Dorta}, {Montesano}, {Muna}, {Myers},
  {Naugle}, {Nichol}, {Noterdaeme}, {Nuza}, {Olmstead}, {Oravetz}, {Oravetz},
  {Owen}, {Padmanabhan}, {Palanque-Delabrouille}, {Pan}, {Parejko},
  {P{\^a}ris}, {Percival}, {P{\'e}rez-Fournon}, {P{\'e}rez-R{\`a}fols},
  {Petitjean}, {Pfaffenberger}, {Pforr}, {Pieri}, {Prada}, {Price-Whelan},
  {Raddick}, {Rebolo}, {Rich}, {Richards}, {Rockosi}, {Roe}, {Ross}, {Ross},
  {Rossi}, {Rubi{\~n}o-Martin}, {Samushia}, {S{\'a}nchez}, {Sayres}, {Schmidt},
  {Schneider}, {Sc{\'o}ccola}, {Seo}, {Shelden}, {Sheldon}, {Shen}, {Shu},
  {Slosar}, {Smee}, {Snedden}, {Stauffer}, {Steele}, {Strauss}, {Streblyanska},
  {Suzuki}, {Swanson}, {Tal}, {Tanaka}, {Thomas}, {Tinker}, {Tojeiro},
  {Tremonti}, {Vargas Maga{\~n}a}, {Verde}, {Viel}, {Wake}, {Watson}, {Weaver},
  {Weinberg}, {Weiner}, {West}, {White}, {Wood-Vasey}, {Yeche}, {Zehavi},
  {Zhao}, \& {Zheng}}]{Dawson2013}
{Dawson}, K.~S., {Schlegel}, D.~J., {Ahn}, C.~P., {et~al.} 2013, \aj, 145, 10

\bibitem[{{de Kool} {et~al.}(2002){de Kool}, {Korista}, \& {Arav}}]{deKool2002}
{de Kool}, M., {Korista}, K.~T., \& {Arav}, N. 2002, \apj, 580, 54

\bibitem[{{Di Matteo} {et~al.}(2005){Di Matteo}, {Springel}, \&
  {Hernquist}}]{DiMatteo2005}
{Di Matteo}, T., {Springel}, V., \& {Hernquist}, L. 2005, \nat, 433, 604

\bibitem[{{Drake} {et~al.}(2009){Drake}, {Djorgovski}, {Mahabal}, {Beshore},
  {Larson}, {Graham}, {Williams}, {Christensen}, {Catelan}, {Boattini},
  {Gibbs}, {Hill}, \& {Kowalski}}]{Drake2009}
{Drake}, A.~J., {Djorgovski}, S.~G., {Mahabal}, A., {et~al.} 2009, \apj, 696,
  870

\bibitem[{{Elvis}(2000)}]{Elvis2000}
{Elvis}, M. 2000, \apj, 545, 63

\bibitem[{{Fabian}(2012)}]{Fabian2012}
{Fabian}, A.~C. 2012, \araa, 50, 455

\bibitem[{{Farrah} {et~al.}(2007){Farrah}, {Lacy}, {Priddey}, {Borys}, \&
  {Afonso}}]{Farrah2007}
{Farrah}, D., {Lacy}, M., {Priddey}, R., {Borys}, C., \& {Afonso}, J. 2007,
  \apjl, 662, L59

\bibitem[{{Filiz Ak} {et~al.}(2013){Filiz Ak}, {Brandt}, {Hall}, {Schneider},
  {Anderson}, {Hamann}, {Lundgren}, {Myers}, {P{\^a}ris}, {Petitjean}, {Ross},
  {Shen}, \& {York}}]{Filiz2013}
{Filiz Ak}, N., {Brandt}, W.~N., {Hall}, P.~B., {et~al.} 2013, \apj, 777, 168

\bibitem[{{Gibson} {et~al.}(2009){Gibson}, {Jiang}, {Brandt}, {Hall}, {Shen},
  {Wu}, {Anderson}, {Schneider}, {Vanden Berk}, {Gallagher}, {Fan}, \&
  {York}}]{Gibson2009a}
{Gibson}, R.~R., {Jiang}, L., {Brandt}, W.~N., {et~al.} 2009, \apj, 692, 758

\bibitem[{{Grier} {et~al.}(2015){Grier}, {Hall}, {Brandt}, {Trump}, {Shen},
  {Vivek}, {Filiz Ak}, {Chen}, {Dawson}, {Denney}, {Green}, {Jiang},
  {Kochanek}, {McGreer}, {P{\^a}ris}, {Peterson}, {Schneider}, {Tao},
  {Wood-Vasey}, {Bizyaev}, {Ge}, {Kinemuchi}, {Oravetz}, {Pan}, \&
  {Simmons}}]{Grier2015}
{Grier}, C.~J., {Hall}, P.~B., {Brandt}, W.~N., {et~al.} 2015, \apj, 806, 111

\bibitem[{{Grier} {et~al.}(2016){Grier}, {Brandt}, {Hall}, {Trump}, {Filiz Ak},
  {Anderson}, {Green}, {Schneider}, {Sun}, {Vivek}, {Beatty}, {Brownstein}, \&
  {Roman-Lopes}}]{Grier2016}
{Grier}, C.~J., {Brandt}, W.~N., {Hall}, P.~B., {et~al.} 2016, \apj, 824, 130

\bibitem[{{Hall} {et~al.}(2011){Hall}, {Anosov}, {White}, {Brandt}, {Gregg},
  {Gibson}, {Becker}, \& {Schneider}}]{Hall2011}
{Hall}, P.~B., {Anosov}, K., {White}, R.~L., {et~al.} 2011, \mnras, 411, 2653

\bibitem[{{Hamann}(1997)}]{Hamann1997}
{Hamann}, F. 1997, \apjs, 109, 279

\bibitem[{{Hamann} {et~al.}(2011){Hamann}, {Kanekar}, {Prochaska}, {Murphy},
  {Ellison}, {Malec}, {Milutinovic}, \& {Ubachs}}]{Hamann2011}
{Hamann}, F., {Kanekar}, N., {Prochaska}, J.~X., {et~al.} 2011, \mnras, 410,
  1957

\bibitem[{{Hamann} {et~al.}(2008){Hamann}, {Kaplan}, {Rodr{\'{\i}}guez
  Hidalgo}, {Prochaska}, \& {Herbert-Fort}}]{Hamann2008}
{Hamann}, F., {Kaplan}, K.~F., {Rodr{\'{\i}}guez Hidalgo}, P., {Prochaska},
  J.~X., \& {Herbert-Fort}, S. 2008, \mnras, 391, L39

\bibitem[{{Hamann} \& {Sabra}(2004)}]{Hamann2004}
{Hamann}, F., \& {Sabra}, B. 2004, in Astronomical Society of the Pacific
  Conference Series, Vol. 311, AGN Physics with the Sloan Digital Sky Survey,
  ed. G.~T. {Richards} \& P.~B. {Hall}, 203

\bibitem[{{Hamann} {et~al.}(2012){Hamann}, {Simon}, {Rodriguez Hidalgo}, \&
  {Capellupo}}]{Hamann2012}
{Hamann}, F., {Simon}, L., {Rodriguez Hidalgo}, P., \& {Capellupo}, D. 2012, in
  Astronomical Society of the Pacific Conference Series, Vol. 460, AGN Winds in
  Charleston, ed. G.~{Chartas}, F.~{Hamann}, \& K.~M. {Leighly}, 47

\bibitem[{{Hamann} {et~al.}(2001){Hamann}, {Barlow}, {Chaffee}, {Foltz}, \&
  {Weymann}}]{Hamann2001}
{Hamann}, F.~W., {Barlow}, T.~A., {Chaffee}, F.~C., {Foltz}, C.~B., \&
  {Weymann}, R.~J. 2001, \apj, 550, 142

\bibitem[{{He} {et~al.}(2017){He}, {Wang}, {Zhou}, {Bian}, {Liu}, {Yang},
  {Dou}, \& {Sun}}]{He2017}
{He}, Z., {Wang}, T., {Zhou}, H., {et~al.} 2017, \apjs, 229, 22

\bibitem[{{Hopkins} {et~al.}(2008){Hopkins}, {Hernquist}, {Cox}, \& {Kere{\v
  s}}}]{Hopkins2008}
{Hopkins}, P.~F., {Hernquist}, L., {Cox}, T.~J., \& {Kere{\v s}}, D. 2008,
  \apjs, 175, 356

\bibitem[{{Leighly} {et~al.}(2009){Leighly}, {Hamann}, {Casebeer}, \&
  {Grupe}}]{Leighly2009}
{Leighly}, K.~M., {Hamann}, F., {Casebeer}, D.~A., \& {Grupe}, D. 2009, \apj,
  701, 176

\bibitem[{{Lu} \& {Lin}(2018{\natexlab{a}})}]{Lu2018complex1}
{Lu}, W.-J., \& {Lin}, Y.-R. 2018{\natexlab{a}}, \mnras, 474, 3397

\bibitem[{{Lu} \& {Lin}(2018{\natexlab{b}})}]{Lu2018complex2}
---. 2018{\natexlab{b}}, \apj, 863, 186

\bibitem[{{Lu} \& {Lin}(2018{\natexlab{c}})}]{Lu2018saturation}
---. 2018{\natexlab{c}}, \apj, 862, 46

\bibitem[{Lu {et~al.}(2018)Lu, Lin, \& Qin}]{LLQ2018}
Lu, W.-J., Lin, Y.-R., \& Qin, Y.-P. 2018, \mnras, 473, L106

\bibitem[{{Lu} {et~al.}(2017){Lu}, {Lin}, {Qin}, {Huang}, {Pan}, {Huang},
  {Yao}, {Nong}, {Yao}, {Lu}, {Pan}, {Huang}, \& {Han}}]{Lu2017}
{Lu}, W.-J., {Lin}, Y.-R., {Qin}, Y.-P., {et~al.} 2017, \mnras, 468, L6

\bibitem[{{McGraw} {et~al.}(2018){McGraw}, {Shields}, {Hamann}, {Capellupo}, \&
  {Herbst}}]{McGraw2018}
{McGraw}, S.~M., {Shields}, J.~C., {Hamann}, F.~W., {Capellupo}, D.~M., \&
  {Herbst}, H. 2018, \mnras, 475, 585

\bibitem[{{McGraw} {et~al.}(2017){McGraw}, {Brandt}, {Grier}, {Filiz Ak},
  {Hall}, {Schneider}, {Anderson}, {Green}, {Hutchinson}, {Macleod}, \&
  {Vivek}}]{McGraw2017}
{McGraw}, S.~M., {Brandt}, W.~N., {Grier}, C.~J., {et~al.} 2017, \mnras, 469,
  3163

\bibitem[{{Misawa} {et~al.}(2014{\natexlab{a}}){Misawa}, {Charlton}, \&
  {Eracleous}}]{Misawa2014a}
{Misawa}, T., {Charlton}, J.~C., \& {Eracleous}, M. 2014{\natexlab{a}}, \apj,
  792, 77

\bibitem[{{Misawa} {et~al.}(2007){Misawa}, {Eracleous}, {Charlton}, {Ganguly},
  {Tytler}, {Kirkman}, {Suzuki}, \& {Lubin}}]{Misawa2007}
{Misawa}, T., {Eracleous}, M., {Charlton}, J.~C., {et~al.} 2007, in
  Astronomical Society of the Pacific Conference Series, Vol. 373, The Central
  Engine of Active Galactic Nuclei, ed. L.~C. {Ho} \& J.-W. {Wang}, 291

\bibitem[{{Misawa} {et~al.}(2005){Misawa}, {Eracleous}, {Charlton}, \&
  {Tajitsu}}]{Misawa2005}
{Misawa}, T., {Eracleous}, M., {Charlton}, J.~C., \& {Tajitsu}, A. 2005, \apj,
  629, 115

\bibitem[{{Misawa} {et~al.}(2014{\natexlab{b}}){Misawa}, {Inada}, {Oguri},
  {Gandhi}, {Horiuchi}, {Koyamada}, \& {Okamoto}}]{Misawa2014b}
{Misawa}, T., {Inada}, N., {Oguri}, M., {et~al.} 2014{\natexlab{b}}, \apjl,
  794, L20

\bibitem[{{Moe} {et~al.}(2009){Moe}, {Arav}, {Bautista}, \&
  {Korista}}]{Moe2009}
{Moe}, M., {Arav}, N., {Bautista}, M.~A., \& {Korista}, K.~T. 2009, \apj, 706,
  525

\bibitem[{{Moravec} {et~al.}(2017){Moravec}, {Hamann}, {Capellupo}, {McGraw},
  {Shields}, \& {Rodr{\'{\i}}guez Hidalgo}}]{Moravec2017}
{Moravec}, E.~A., {Hamann}, F., {Capellupo}, D.~M., {et~al.} 2017, \mnras, 468,
  4539

\bibitem[{{Murray} \& {Chiang}(1995)}]{Murray1995}
{Murray}, N., \& {Chiang}, J. 1995, \apjl, 454, L105

\bibitem[{{Narayanan} {et~al.}(2004){Narayanan}, {Hamann}, {Barlow},
  {Burbidge}, {Cohen}, {Junkkarinen}, \& {Lyons}}]{Narayanan2004}
{Narayanan}, D., {Hamann}, F., {Barlow}, T., {et~al.} 2004, \apj, 601, 715

\bibitem[{{Osterbrock} \& {Ferland}(2006)}]{Osterbrock2006}
{Osterbrock}, D.~E., \& {Ferland}, G.~J. 2006, {Astrophysics of gaseous nebulae
  and active galactic nuclei}

\bibitem[{{P{\^a}ris} {et~al.}(2017){P{\^a}ris}, {Petitjean}, {Ross}, {Myers},
  {Aubourg}, {Streblyanska}, {Bailey}, {Armengaud}, {Palanque-Delabrouille},
  {Y{\`e}che}, {Hamann}, {Strauss}, {Albareti}, {Bovy}, {Bizyaev}, {Niel
  Brandt}, {Brusa}, {Buchner}, {Comparat}, {Croft}, {Dwelly}, {Fan},
  {Font-Ribera}, {Ge}, {Georgakakis}, {Hall}, {Jiang}, {Kinemuchi},
  {Malanushenko}, {Malanushenko}, {McMahon}, {Menzel}, {Merloni}, {Nandra},
  {Noterdaeme}, {Oravetz}, {Pan}, {Pieri}, {Prada}, {Salvato}, {Schlegel},
  {Schneider}, {Simmons}, {Viel}, {Weinberg}, \& {Zhu}}]{Paris2017}
{P{\^a}ris}, I., {Petitjean}, P., {Ross}, N.~P., {et~al.} 2017, \aap, 597, A79

\bibitem[{{Proga} {et~al.}(2000){Proga}, {Stone}, \& {Kallman}}]{Proga2000}
{Proga}, D., {Stone}, J.~M., \& {Kallman}, T.~R. 2000, \apj, 543, 686

\bibitem[{{Rodr{\'{\i}}guez Hidalgo} {et~al.}(2013){Rodr{\'{\i}}guez Hidalgo},
  {Eracleous}, {Charlton}, {Hamann}, {Murphy}, \& {Nestor}}]{Rodriguez2013}
{Rodr{\'{\i}}guez Hidalgo}, P., {Eracleous}, M., {Charlton}, J., {et~al.} 2013,
  \apj, 775, 14

\bibitem[{{Rogerson} {et~al.}(2016){Rogerson}, {Hall}, {Rodr{\'{\i}}guez
  Hidalgo}, {Pirkola}, {Brandt}, \& {Filiz Ak}}]{Rogerson2016}
{Rogerson}, J.~A., {Hall}, P.~B., {Rodr{\'{\i}}guez Hidalgo}, P., {et~al.}
  2016, \mnras, 457, 405

\bibitem[{{Sabra} \& {Hamann}(2005)}]{Sabra2005}
{Sabra}, B.~M., \& {Hamann}, F. 2005, ArXiv Astrophysics e-prints,
  astro-ph/0509421

\bibitem[{{Silk} \& {Rees}(1998)}]{Silk1998}
{Silk}, J., \& {Rees}, M.~J. 1998, \aap, 331, L1

\bibitem[{{Stathopoulos} {et~al.}(2019){Stathopoulos}, {Danezis}, {Lyratzi},
  {Antoniou}, \& {Tzimeas}}]{Stathopoulos2019}
{Stathopoulos}, D., {Danezis}, E., {Lyratzi}, E., {Antoniou}, A., \& {Tzimeas},
  D. 2019, \mnras, 486, 894

\bibitem[{{Vivek}(2019)}]{Vivek2019}
{Vivek}, M. 2019, \mnras, 486, 2379

\bibitem[{{Wang} {et~al.}(2015){Wang}, {Yang}, {Wang}, \& {Ferland}}]{Wang2015}
{Wang}, T., {Yang}, C., {Wang}, H., \& {Ferland}, G. 2015, \apj, 814, 150

\bibitem[{{Weymann} {et~al.}(1991){Weymann}, {Morris}, {Foltz}, \&
  {Hewett}}]{Weymann1991}
{Weymann}, R.~J., {Morris}, S.~L., {Foltz}, C.~B., \& {Hewett}, P.~C. 1991,
  \apj, 373, 23

\bibitem[{{York} {et~al.}(2000){York}, {Adelman}, {Anderson}, {Anderson},
  {Annis}, {Bahcall}, {Bakken}, {Barkhouser}, {Bastian}, {Berman}, {Boroski},
  {Bracker}, {Briegel}, {Briggs}, {Brinkmann}, {Brunner}, {Burles}, {Carey},
  {Carr}, {Castander}, {Chen}, {Colestock}, {Connolly}, {Crocker}, {Csabai},
  {Czarapata}, {Davis}, {Doi}, {Dombeck}, {Eisenstein}, {Ellman}, {Elms},
  {Evans}, {Fan}, {Federwitz}, {Fiscelli}, {Friedman}, {Frieman}, {Fukugita},
  {Gillespie}, {Gunn}, {Gurbani}, {de Haas}, {Haldeman}, {Harris}, {Hayes},
  {Heckman}, {Hennessy}, {Hindsley}, {Holm}, {Holmgren}, {Huang}, {Hull},
  {Husby}, {Ichikawa}, {Ichikawa}, {Ivezi{\'c}}, {Kent}, {Kim}, {Kinney},
  {Klaene}, {Kleinman}, {Kleinman}, {Knapp}, {Korienek}, {Kron}, {Kunszt},
  {Lamb}, {Lee}, {Leger}, {Limmongkol}, {Lindenmeyer}, {Long}, {Loomis},
  {Loveday}, {Lucinio}, {Lupton}, {MacKinnon}, {Mannery}, {Mantsch}, {Margon},
  {McGehee}, {McKay}, {Meiksin}, {Merelli}, {Monet}, {Munn}, {Narayanan},
  {Nash}, {Neilsen}, {Neswold}, {Newberg}, {Nichol}, {Nicinski}, {Nonino},
  {Okada}, {Okamura}, {Ostriker}, {Owen}, {Pauls}, {Peoples}, {Peterson},
  {Petravick}, {Pier}, {Pope}, {Pordes}, {Prosapio}, {Rechenmacher}, {Quinn},
  {Richards}, {Richmond}, {Rivetta}, {Rockosi}, {Ruthmansdorfer}, {Sandford},
  {Schlegel}, {Schneider}, {Sekiguchi}, {Sergey}, {Shimasaku}, {Siegmund},
  {Smee}, {Smith}, {Snedden}, {Stone}, {Stoughton}, {Strauss}, {Stubbs},
  {SubbaRao}, {Szalay}, {Szapudi}, {Szokoly}, {Thakar}, {Tremonti}, {Tucker},
  {Uomoto}, {Vanden Berk}, {Vogeley}, {Waddell}, {Wang}, {Watanabe},
  {Weinberg}, {Yanny}, {Yasuda}, \& {SDSS Collaboration}}]{York2000}
{York}, D.~G., {Adelman}, J., {Anderson}, Jr., J.~E., {et~al.} 2000, \aj, 120,
  1579

\end{thebibliography}

\end{document}